\documentclass[preprint,12pt,3p]{elsarticle}

\usepackage{amssymb}
\usepackage{amsmath}

\journal{Physics Letters A}

\begin{document}

\begin{frontmatter}

\title{Stability boundaries of a Mathieu equation having $PT$ symmetry}

\author{P. A. Brand\~ao}
\address{Instituto de F\'isica, Universidade Federal de Alagoas, Macei\'o - AL}


\ead{paulo.brandao@fis.ufal.br}



\begin{abstract}
I have applied multiple-scale perturbation theory to a generalized complex $PT$-symmetric Mathieu equation in order to find the stability boundaries between bounded and unbounded solutions. The analysis suggests that the non-Hermitian parameter present in the equation can be used to control the shape and curvature of these boundaries. Although this was suggested earlier by several authors, analytic formulas for the boundary curves were not given. This paper is a first attempt to fill this gap in the theory.
\end{abstract}


\end{frontmatter}


\section{Introduction}

Non-Hermitian Hamiltonians having $PT$-symmetry were introduced in 1998 by Bender and Boettcher in a first attempt to generalize the standard postulates of Hermitian quantum mechanics to include operators invariant under parity ($\mathcal{P}$) and time reversal ($\mathcal{T}$) symmetries \cite{bender1}. The authors showed that if a Hamiltonian commutes with the $PT$ operator, the energy eigenvalues could consist of entirely real, discrete and positive numbers. However, this is only a necessary but not a sufficient condition for the reality of the spectrum. If a $PT$-symmetric Hamiltonian is found to possess complex eigenvalues, we say that it has a broken symmetric phase \cite{bender2,bender3,bender4}. If a $PT$-symmetric Hamiltonian is found to possess real spectra, then it is possible to introduce a new symmetry, described by an operator $\mathcal{C}$, such that the evolution of the entire system is unitary, consistent with physical requirements of probability conservation \cite{bender3}. The condition imposed on the potential function $V(x)$ of the Schr\"odinger equation to guarantee $PT$ symmetry is that $V(x) = V(-x)^{*}$, as can be easily verified. This, in turn, implies that the real (imaginary) part of the potential must be even (odd) under $x\rightarrow -x$. Reviews on the subject of $PT$ symmetry in quantum and classical mechanics are already available \cite{bender5,bender6}. 

The Schr\"odinger equation with a periodic potential $V(x) = V(x+P)$ ($P$ being the period) having parity and time-reversal symmetry was initially considered by Bender and coauthors \cite{bender7} and also by Z. Ahmed \cite{ahmed}. The authors found that real energy bands can be created even though the potential is a complex function of position. Moreover, the bands could be controlled by tuning the non-Hermitian parameter present in the Hamiltonian. There is also a vast literature dealing with the dynamics of wavepackets propagating in periodic potentials with $PT$ symmetry \cite{makris2008,mei2010,makris2010,guo2009,avinash2004,longhibloch,brandao2017}. Apart from a purely mathematical context, the Mathieu equation is one of the most studied second order linear differential equation in applied mathematics \cite{mathieu1,mathieu2}. Its solutions, known as Mathieu functions, play a key role in physics with special interest in the propagation of electrons in crystal lattices, where the potential function of the Schr\"odinger equation reduces to a Mathieu type. The Mathieu equation is a special case of a more general second order linear differential equation known as a Hill's equation \cite{hillbook}. Given a differential equation of a Mathieu type,
\begin{equation}\label{mathieu}
    \frac{d^2 \psi(x)}{dx^2} + (a + 2\varepsilon \cos x)\psi(x) = 0,
\end{equation}
where $a$ and $\varepsilon$ are real and positive constants, it is well known that, depending on the numerical values of $(a,\varepsilon)$, the solutions can be bounded or unbounded with respect to $x$. One of the main objectives is to find the boundary lines in the $(a,\varepsilon)$-plane which separates bounded from unbounded solutions. The problem of finding the stability boundaries of a Mathieu equation is well documented \cite{bookbender}. The fact that one can control the band diagram by tuning the non-Hermitian character of the lattice is also not a new result and it was suggested earlier by several authors \cite{bender7,makris2008,makris2010,yang2012,ahmed2001}. There are also some very interesting connections between the Mathieu equation and the E2 algebra in non-Hermitian models \cite{kalveks2011,fring2015,fring20152}. However, to my knowledge, a detailed analysis aimed to obtain analytic expressions for the stability boundaries in the $(a,\varepsilon)$ plane for a $PT$ symmetric lattice has not been considered. Therefore, the objective of this paper is to close this gap by deriving the first set of approximate expressions that can be used to understand in a clear way the shape and curvature of the stability boundaries of a generalized Mathieu equation with $PT$ symmetry.

\section{Multiple-Scale Analysis: Setting up the problem}

The primary goal is to find the boundaries between stable and unstable solutions in the $(a,\varepsilon)$-plane of the generalized $PT$-symmetric Mathieu equation and to understand their behavior as the non-Hermitian parameter $\beta\in [0,1]$ varies. The choice for this range of $\beta$ is dictated by the fact that when $\beta > 1$ the Schr\"odinger equation is no longer normalizable under a $CPT$ inner product and therefore it is not possible to construct a physical theory for values outside this range \cite{bender3}. In other words, the Hamiltonian has a broken phase when $\beta > 1$ \cite{betaone}. To this end, define the $PT$-symmetric Mathieu equation
\begin{equation}
\label{mat}
    \frac{d^2 \psi(x)}{dx^2} + [a + 2\varepsilon V(x)]\psi(x) = 0,
\end{equation}
where the periodic potential $V(x) = \cos x + i\beta\sin x$ has $PT$ symmetry since it satisfies $V(x) = V(-x)^{*}$ as discussed previously. If $\beta = 0$, the usual Mathieu equation is recovered, whose solutions converge for $a_n \neq n^2/4$ ($n = 0,1,2,...$ and $\varepsilon\rightarrow 0$), as can be easily verified by using regular perturbation expansions around small values of $\varepsilon$ \cite{bookbender}. Multiple-scale perturbation analysis starts by assuming that the solution can be expanded in a perturbation series of the form
\begin{equation}
\label{psi}
    \psi(x,\tau) = \sum_{n=0}^{\infty}\varepsilon^n Y_n(x,\tau) = Y_0(x,\tau) + \varepsilon Y_1(x,\tau) + \varepsilon^2 Y_2(x,\tau) + ...
\end{equation}
where $\tau = \varepsilon x$ is a long spatial scale (treated as an independent variable) since $\tau$ is not negligible when $x$ is of order $1/\varepsilon$ or larger. By taking the total second order derivative of \eqref{psi} with respect to $x$, we obtain the expansion
\begin{align}
\label{exp}
    \frac{d^2 \psi}{dx^2} =  \sum_{n=0}^{\infty}\varepsilon^{n}\left( \frac{\partial^2 Y_n}{\partial x^2} + 2\varepsilon\frac{\partial^2 Y_n}{\partial\tau \partial x} + \varepsilon^2\frac{\partial^2 Y_n}{\partial \tau^2} \right)
\end{align}
We also expand the parameter $a$ in a perturbation series in powers of $\varepsilon$: 
\begin{equation}
\label{aexp}
a(\varepsilon) = \sum_{n=0}^{\infty}a_n\varepsilon^n  = a_0 + a_1\varepsilon + a_2\varepsilon^2 + ... 
\end{equation}
where $a_j$ are real numbers. It should be mentioned that I am assuming that expansions \eqref{psi} and \eqref{aexp} are a good representation of the true solution at least as long as $\varepsilon\rightarrow 0$. Since multiple-scale perturbation performs a global analysis, we expect the results to be valid at every point on the $x$ axis. In general, it is only through numerical methods applied to Equation \eqref{mat} that we can really obtain any sort of information about the errors in the expansion \cite{bookbender}. However, since my objective is to obtain analytical formulas for the stability boundaries, numerical methods are usually not enlightening in this matter and that is why I chose to work with perturbation expansions. After substituting \eqref{psi}, \eqref{exp} and \eqref{aexp} into \eqref{mat}, and collecting powers of $\varepsilon$, we obtain the following system of coupled linear partial differential equations:
\begin{align}
\label{sys1}
    \varepsilon^{0}: \frac{\partial^2 Y_0}{\partial x^2} + a_0 Y_0 &= 0, \\
        \label{sys2}
    \varepsilon^{1}: \frac{\partial^2 Y_1}{\partial x^2} + a_0 Y_1 &= -a_1 Y_0 - 2V(x)Y_0 -2\frac{\partial^2 Y_0}{\partial \tau \partial x}, \\
    \label{sys3}
    \varepsilon^{n}: \frac{\partial^2 Y_n}{\partial x^2} + a_0 Y_n &= -\sum_{k=1}^{n}a_k Y_{n-k} - 2V(x)Y_{n-1} -  2\frac{\partial^2 Y_{n-1}}{\partial \tau \partial x} - \frac{\partial^2 Y_{n-2}}{\partial \tau^2} \hspace{0.5cm} (n\geq 2).
\end{align}
Equations \eqref{sys1}, \eqref{sys2} and \eqref{sys3} provide the starting point for the analysis. In what follows I am going to consider the expansion around two particular values of $a_{0}$: \{0, $\frac{1}{4}$\} in order to find the stability boundaries between stable and unstable solutions in the vicinity of these specific values. As I pointed out at the beginning of this section, the solutions are stable if $a_n \neq n^2/4$ ($n = 0, 1, 2,...$) in the vicinity of $\varepsilon\rightarrow 0$ and therefore the stability lines cut the $a$-axis at these points. The analysis presented here can be easily applied to any value of $a_0$ and the above two values are considered because the formulas for the stability lines around $0$ and $1/4$ are already known (for the Hermitian case) and so we can compare the results obtained here with the literature \cite{bookbender}.

\section{Perturbative analysis for $a_0 = 0$}

The structure of the set of coupled equations \eqref{sys1}-\eqref{sys3} is such that if $Y_0$ is found from \eqref{sys1}, $Y_1$ can be calculated from \eqref{sys2} and all the successive orders can be found from \eqref{sys3} where the right hand side depends on previous calculated values. Therefore, we first set $a_0 = 0$ in \eqref{sys1} such that the function $Y_0$ satisfies 
\begin{equation}
\frac{\partial^2 Y_0}{\partial x^2} = 0     
\end{equation}
and the bounded solution is $Y_0(x,\tau) = h(\tau)$ with $h(\tau)$ an arbitrary bounded function of $\tau$. To find $a_1$ we substitute the above $Y_0$ into \eqref{sys2} to obtain
\begin{equation}
\label{Y1}
    \frac{\partial^2 Y_1}{\partial x^2} = -h(\tau)[a_1 + 2(\cos x + i\beta \sin x)].
\end{equation}
If we impose the condition that $Y_1$ be a bounded function, we must take $a_1 = 0$. Therefore, we have the $a$ expansion up to second order in $\varepsilon$: $a = a_2\varepsilon^2$. To find $a_2$ we need the solution of the differential equation \eqref{Y1} with $a_1 = 0$. This is easily done and the result is given by
\begin{equation}
    Y_1 = 2h(\tau)(\cos x + i\beta\sin x) + g(\tau),
\end{equation}
where $g(\tau)$ is an arbitrary bounded function of $\tau$. The differential equation \eqref{sys3} satisfied by $Y_2$ can be written as
\begin{align}
\label{Y2}
    \frac{\partial^2 Y_2}{\partial x^2} &= - h''(\tau) - h(\tau)[a_2  + 2(1-\beta^2)] - 2h(\tau)(1+\beta^2)\cos2x \\
    &- 4ih(\tau)\beta\sin2x -2[g(\tau) + 2i\beta h'(\tau)]\cos x - 2[ig(\tau)\beta - 2h'(\tau)]\sin x.
\end{align}
If we impose the condition that $Y_2$ remain bounded, we must have
\begin{equation}
    h''(\tau) + [a_2 + 2(1-\beta^2)]h(\tau) = 0
\end{equation}
so that
\begin{align}
    a_2 &> -2(1-\beta^2) \hspace{0.5cm} \text{(stable)} \\
    a_2 &< -2(1-\beta^2) \hspace{0.5cm} \text{(unstable)}
\end{align}
and we can take
\begin{equation}
    a_2 = -2(1-\beta^2)
\end{equation}
as the boundary stability line with $h(\tau) = h_0$ a constant value. This gives us the second order contribution to $a$:
\begin{equation}
\label{a0}
    a(\varepsilon) = -2(1-\beta^2)\varepsilon^2 + O(\varepsilon^3),
\end{equation}
so that the curvature $\kappa$ of the boundary is
\begin{equation}
\label{curv1}
    \kappa_1 = \left| \frac{d^2 a(0)}{d\varepsilon^2} \right| = 4(1-\beta^2)
\end{equation}
and the function $Y_2$ (which is $PT$-symmetric if we take $g$ as a real function of $\tau$) is given by
\begin{equation}
    Y_2 = \frac{h_0 (1+\beta^2)}{2}\cos2x + i\beta h_0 \sin2x + 2g(\tau)\cos x + 2ig\beta(\tau)\sin x + f(\tau),
\end{equation}
where $f(\tau)$ is an arbitrary bounded function of $\tau$. This process can be repeated if one wishes to find the third order correction $a_3$ and so on \footnote{It can be shown that $a_3$ is, in fact, zero.}. It is seen from \eqref{a0} and \eqref{curv1} that the non-Hermitian parameter $\beta$ quadratically controls the curvature of the stability boundary for $\varepsilon$ near zero. In particular, if $\beta$ assumes the unit value, which is the critical number in which the symmetry of the lattice is spontaneously broken, the curvature vanishes. Figure 1 shows a plot of the stability line for three values of $\beta$ in the plane $(a,\varepsilon)$. When $\beta = 0$ we recover the usual approximation found in textbooks that $a = -2\varepsilon^2$, shown as a continuous line in the figure. As soon as $\beta\neq 0$ the stability line deforms in such a way that it approximates to a straight line when $\beta$ is close to 1. The introduction of the non-Hermitian parameter therefore has a very important consequence on the position in the $(a,\varepsilon)$-plane of stable and unstable solutions.

\begin{figure}[h]
\centering
\includegraphics[width=0.4\textwidth]{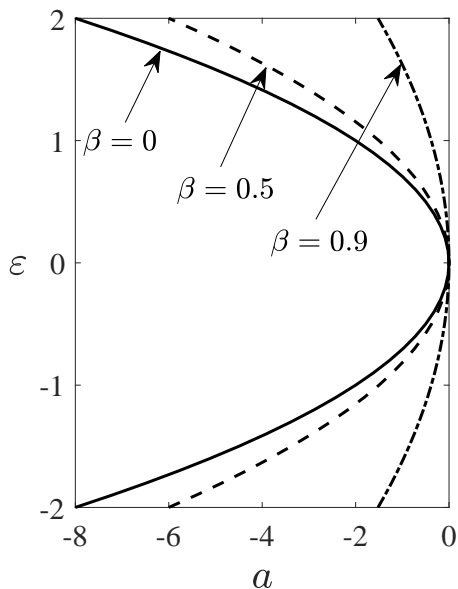}
\caption{Stability boundaries of the $PT$-symmetric Mathieu equation \eqref{mat} (up to second order in $\varepsilon$) in the $(a,\varepsilon)$-plane for $a_0 = 0$ given by equation \eqref{a0}. Three boundaries are shown for $\beta = 0$, $0.5$ and $0.9$. }
\end{figure}

\section{Perturbative analysis for $a_0 = 1/4$}

Let us now consider a slightly more complicated situation. If we choose $a_0 = \frac{1}{4}$ then, according to \eqref{sys1}, the function $Y_0$ satisfies
\begin{equation}
    \frac{\partial^2 Y_0}{\partial x^2} + \frac{Y_0}{4} = 0.
\end{equation}
The general solution of this linear partial differential equation is
\begin{equation}
    Y_0 = A_1 (\tau)e^{ix/2} + A_2(\tau)e^{-ix/2}
\end{equation}
with $A_1$ and $A_2$ arbitrary bounded functions of the parameter $\tau$. According to \eqref{sys2}, the function $Y_1$ satisfies the differential equation
\begin{equation}
\begin{split}
\frac{\partial^2 Y_1}{\partial x^2} + \frac{Y_1}{4} & = [-a_1 A_1 - iA_1 ' - (1+\beta)A_2]e^{ix/2} \\
 & + [-a_1 A_2 + iA_2 ' - (1-\beta)A_1]e^{-ix/2} \\
 & + -(1+\beta)A_1e^{3ix/2} - (1-\beta)A_2 e^{-3ix/2}
\end{split}
\end{equation}
In order to obtain bounded solutions, the secular terms must vanish. This implies the following coupled second order differential equations for $A_1$ and $A_2$:
\begin{align}
    a_1 A_1 + iA_1 ' + (1+\beta)A_2 &= 0, \\
    a_1 A_2 - iA_2 ' + (1-\beta)A_1 &= 0.
\end{align}
Both, $A_1$ and $A_2$, satisfy the same differential equation, namely
\begin{equation}
    A_j '' = [(1-\beta^2) - a_1^2]A_j,
\end{equation}
and it is easy to see that we have two conditions on the boundness of $A_j$:
\begin{align}
    a_1^2 &> 1-\beta^2 \longrightarrow \text{Bounded solutions} \\
    a_1^2 &< 1-\beta^2 \longrightarrow \text{Unbounded solutions}
\end{align}
so that the stability boundary line in the $(a,\varepsilon)$ plane (up to first order in $\varepsilon$) is given by
\begin{equation}
    a(\varepsilon) = \frac{1}{4} \pm \sqrt{1-\beta^2}\varepsilon.
\end{equation}
To find the second order correction, $a_2$, we notice that $Y_1$ is given by
\begin{equation}
\label{Y12}
    Y_1 = \frac{(1+\beta)}{2}K_{+}e^{3ix/2} + \frac{(1-\beta)}{2}K_{+}e^{-3ix/2} + B_1(\tau)e^{ix/2} + B_2(\tau)e^{-ix/2},
\end{equation}
where $B_j$ are arbitrary bounded functions of the parameter $\tau$. After substituting \eqref{Y12} into \eqref{sys3} we get
\begin{align}
    \frac{\partial^2 Y_2}{\partial x^2} + \frac{Y_2}{4} &= e^{ix/2}\left[ -\sqrt{1-\beta^2}B_1 - a_2 K_{+} - iB_1 ' - (1+\beta)B_2 - \frac{(1-\beta^2)}{2}K_{+} \right] \nonumber \\
    &+ e^{-ix/2}\left[ -\sqrt{1-\beta^2}B_2 -a_2 K_{-} + iB_2 ' -\frac{(1-\beta^2)}{2}K_{-} - (1-\beta)B_1 \right] \nonumber \\
    &+ e^{3ix/2}\left[ -\frac{\sqrt{1-\beta^2}(1+\beta)}{2}K_{+} - (1+\beta)B_1\right] \nonumber \\
    &+ e^{-3ix/2}\left[ -\frac{\sqrt{1-\beta^2}(1-\beta)}{2}K_{-} - (1-\beta)B_2 \right] \nonumber\\
    &-\frac{(1+\beta^2)^2}{2}K_{+}e^{7ix/2} - \frac{(1-\beta)^2}{2}K_{-}e^{-5ix/2},
\end{align}
and imposing the condition that the secular terms vanish we have
\begin{equation}
    iB_1 ' + \sqrt{1-\beta^2}B_1 + (1+\beta)B_2 + K_{+}\left[ a_2 + \frac{1-\beta^2}{2} \right] = 0
\end{equation}
and
\begin{equation}
    iB_2 ' + \sqrt{1-\beta^2}B_2 + (1+\beta)B_1 + K_{+}\left[ a_2 + \frac{1-\beta^2}{2} \right] = 0
\end{equation}
which implies that $B_j$ with $j = 1,2$ both satisfy the same differential equation
\begin{equation}
    B_j '' + [ \sqrt{1-\beta^2}K_{+} - (1+\beta)K_{-} ]\left[ a_2 + \frac{(1-\beta^2)}{2} \right] = 0.
\end{equation}
For a bounded solution to exist, we must impose that
\begin{equation}
\label{a22}
    a_2 = -\frac{(1-\beta^2)}{2}
\end{equation}
with $B_j$ a constant independent of $\tau$. Equation \eqref{a22} is consistent with the literature when $\beta = 0$ \cite{bookbender}. Therefore, we can write the boundary line up to second order in $\varepsilon$ as
\begin{equation}
\label{a3}
    a(\varepsilon) = \frac{1}{4} \pm \sqrt{1-\beta^2}\varepsilon -\frac{(1-\beta^2)}{2}\varepsilon^2.
\end{equation}
The curvature $\kappa_2$ of this stability line is now given by 
\begin{equation}
    \kappa_2 = 1-\beta^2 = \frac{\kappa_1}{4},
\end{equation}
which is always less than the curvature of the stability line around $a_0 = 0$. Figure 2 shows the stability lines as given by \eqref{a3}. One can clearly see from the figure that the boundaries can be tuned depending on $\beta$. Once this non-Hermitian parameter acquires physical interpretation (for example, it can be considered as the physical parameter which controls loss and gain in waveguide arrays in optical systems), the above approximated equations for the stability boundaries can be used to engineer a specific function on the Bloch waves. It should be noted that these lines are a global property of the system, implying that Figures 1 and 2 are composing the same system for the same values of $\beta$.
\begin{figure}[h]
\centering
\includegraphics[width=0.4\textwidth]{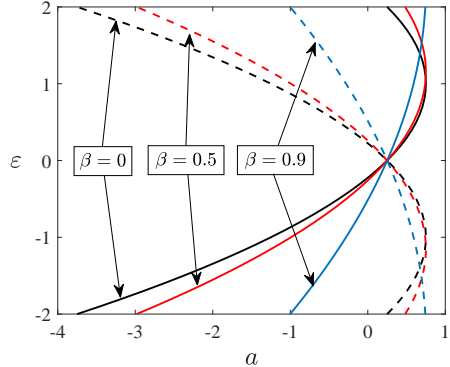}
\caption{(Color online) Stability boundaries of a $PT$-symmetric Mathieu equation in the $(a,\varepsilon)$ plane valid for $a$ near $\frac{1}{4}$. The continuous (dashed) lines are obtained from Equation \eqref{a3} with the plus (minus) sign. Black lines indicate the Hermitian case where $\beta = 0$ and red and blue lines are non-Hermitian cases with $\beta = 0.5$ and $\beta = 0.9$, respectively.  }
\end{figure}

\section{Conclusions}

The introduction of a non-Hermitian term into the Mathieu equation \eqref{mathieu} [giving a generalized complex Mathieu equation \eqref{mat}] opens new possibilities in the exploration of the band structure of physical systems. It is well known that the Schr\"odinger equation with a periodic potential is the main tool behind solid state physics, and, within the independent electron approximation, the band structure of the material determines, for example, its conductivity and optical properties. Therefore, it is very important to understand the topological aspects of the band structure and we analytically provide the first set of equations that may be used to characterize the physical material [see Equations \eqref{a0} and \eqref{a3}]. Although this result was verified numerically by a number of authors, the exact form of these stability lines were not known. I have also shown that the curvature of these lines (as $\varepsilon\rightarrow 0$) are related to the non-Hermitian parameter $\beta$. Thus, in principle, if one can measure the curvatures in some experimental arrangement, the determination of $\beta$ could be obtained in laboratory conditions. Future work will be aimed at the deformation of the stability lines as the Mathieu equation becomes nonlinear and non-Hermitian. These results will be published in near future.

\section*{Acknowledgements}
The author would like to acknowledge the Brazilian
Agencies CNPq, CAPES and FAPEAL for partial financial support.




\bibliographystyle{elsarticle-num}


\begin{thebibliography}{99}
\bibitem{bender1} 
C. M. Bender, and S. Boettcher, ``Real spectra in non-Hermitian Hamiltonians having $PT$ symmetry,'' Phys. Rev. Lett. \textbf{80}, 5243 (1998).

\bibitem{bender2}
C. M. Bender, S. Boettcher, and P. N. Meisinger, ``PT-symmetric quantum mechanics,'' J. Math. Phys. \textbf{40}, 2201 (1999).

\bibitem{bender3}
C. M. Bender, C. D. Brody, and H. F. Jones, ``Complex Extension of Quantum Mechanics,'' Phys. Rev. Lett. \textbf{89}, 270401 (2002).

\bibitem{bender4}
C. M. Bender, C. D. Brody, and H. F. Jones, ``Must a Hamiltonian be Hermitian?'', Am. J. Phys. \textbf{71}, 1095-1102 (2003).

\bibitem{bender5}
C. M. Bender, ``Introduction to PT-symmetric quantum mechanics'', Cont. Phys. \textbf{46}, 277-292 (2005).

\bibitem{bender6} 
C. M. Bender, ``Making sense of non-Hermitian Hamiltonians'', Rep. Prog. Phys. \textbf{70}, 947-1018 (2007).

\bibitem{bender7}
C. M. Bender, ``Complex periodic potentials with real band spectra'', Phys, Lett. A \textbf{252}, 272-276 (1999).

\bibitem{ahmed} Z. Ahmed, ``Energy band structure due to a complex, periodic, $\mathcal{PT}$-invariant potential'', Phys. Lett. A, \textbf{286}, 231 (2001).

\bibitem{makris2008} K. G. Makris, R. El-Ganainy, D. N. Christodoulides, and Z. H. Musslimani, ``Beam Dynamics in $PT$ Symmetric Optical Lattices'', Phys. Rev. Lett. \textbf{103904}, 103904 (2008).

\bibitem{mei2010} Mei C. Zheng, Demetrios N. Christodoulides, Ragnar Fleischmann, and Tsampikos Kottos, ``$\mathcal{PT}$ optical lattices and universality in beam dynamics'', Phys. Rev. A \textbf{82}, 010103(R) (2010).

\bibitem{makris2010} Konstantinos G. Makris, Ramy El-Ganainy, Demetrios N. Christodoulides, and Z. H. Musslimani, ``$\mathcal{PT}$-symmetric optical lattices'', Phys. Rev. A, \textbf{81}, 063807 (2010).

\bibitem{guo2009} A. Guo, G. J. Salamo, D. Duchesne, R. Morandotti, M. Volatier-Ravat, V. Aimez, G. A. Siviloglou, and D. N. Christodoulides, ``Observation of $\mathcal{PT}$-Symmetry Breaking in Complex Optical Potentials'', Phys. Rev. Lett., \textbf{103}, 093902 (2009).

\bibitem{avinash2004} A. Khare and Uday Sukhatme, ``Analytically solvable $\mathcal{PT}$-invariant periodic potentials'', Phys. Lett. A, \textbf{324}, 406 (2004).

\bibitem{longhibloch} S. Longhi, ``Bloch Oscillations in Complex Crystals with $\mathcal{PT}$ Symmetry'', Phys. Rev. Lett. \textbf{103}, 123601 (2009).

\bibitem{brandao2017} P. A. Brand\~ao and S. B. Cavalcanti, ``Bragg-induced power oscillations in $\mathcal{PT}$-symmetric periodic photonic structures'', Phys. Rev. A, \textbf{96},  053841 (2017).

\bibitem{mathieu1}
N. W. McLachlan, ``Theory and application of Mathieu functions'' (Dover, 1964).

\bibitem{mathieu2}
F. M. Arscott, ``Periodic Differential Equations: An Introduction to Mathieu, Lam\'e, and Allied Functions'' (Pergamon, 2014).

\bibitem{hillbook}
W. Magnus, and S. Winkler, ``Hill's equation'' (Dover, 1979).

\bibitem{bookbender}
C. M. Bender, and S. A. Orszag, \text{Advanced mathematical methods for scientist and engineers} (Springer, 1999).

\bibitem{yang2012} S. Nixon, L. Ge, and J. Yang, ``Stability analysis for solitons in PT-symmetric optical lattices'', Phys. Rev. A \textbf{85}, 023822 (2012).

\bibitem{ahmed2001} Z. Ahmed. ``Energy band structure due to a complex, periodic, PT-invariant potential'', Phys. Lett. A \textbf{286}, 231 (2001).

\bibitem{kalveks2011} Carl M. Bender and J. Kalveks, ``Extending $\mathcal{PT}$ Symmetry from Heisenberg Algebra to E2 Algebra'', Int. J. Theor. Phys. \textbf{50}, 955 (2011). 

\bibitem{fring2015} A. Fring, ``E2-quasi-exact solvability for non-Hermitian models'', J. Phys. A: Math. Theor. \textbf{48}, 145301 (2015). 

\bibitem{fring20152} A. Fring, ``A new non-Hermitian E2-quasi-exactly solvable model'', Phys. Lett. A, \textbf{379}, 873 (2015).

\bibitem{betaone} It should be noted that the regime $\beta > 1$ has interesting applications in the optical context of the paraxial wave equation with $PT$ symmetry (see refs. \cite{makris2008,mei2010,makris2010,guo2009,avinash2004,longhibloch,brandao2017}).


\end{thebibliography}

\end{document}